\documentstyle{article}[11pt]
\newcommand{\ltsimeq}{\raisebox{-0.6ex}{$\,\stackrel
{\raisebox{-.2ex}{$\textstyle <$}}{\sim}\,$}}

\begin{document}
\begin{center}
\Large{A High Resolution Radio Survey of Class I Protostars}

\vspace{6mm}
\large{P.W.Lucas$^{1,2}$, Katherine M.Blundell$^{2}$ and P.F.Roche$^{2}$}
\end{center}

\vspace{4mm}
$^1$Dept. of Physical Sciences, University of Hertfordshire, College Lane,
Hatfield AL10 9AB.\\ email: pwl@star.herts.ac.uk\\

$^2$Astrophysics Dept., University of Oxford, 1 Keble Road, Oxford OX1 3RH.

\vspace{4mm}
Accepted by MNRAS

\vspace{1cm}
\large{\textbf{Abstract}}
\normalsize

	We report the results of a survey of low mass Class I protostars
in the cm continuum. In the initial survey, seven sources in the Taurus 
star formation were observed with the VLA at 0$^{``}$.25 resolution. All 
seven sources drive CO outflows and display Herbig-Haro 
flows in the optical or near infrared wavebands. 4/7 sources were detected, 
two of which are new discoveries in systems of very low luminosity, one being 
the lowest luminosity system detected to date in the cm continuum. Notably, 
three sources were not detected to a 3-$\sigma$ limit of 0.10~mJy/beam, which 
indicates that significant cm continuum emission is not a universal feature 
of Class I systems with outflow activity. Subsequent observations of HH30, a 
more evolved Class II system, found no emission to a 3-$\sigma$ limit of 
0.03~mJy/beam. After 
comparison with near infrared data, we suggest that the discriminating feature 
of the detected systems is a relatively high ionisation fraction in the 
stellar wind. Temporal variability of the outflow may also play a role: only 
recently ejected knots may have sufficiently dense plasma to be optically 
thick to free-free emission, and hence produce detectable flux. The one 
relatively bright source, IRAS 04016+2610 (L1489 IRS), is clearly resolved on 
a 0$^{``}$.4 scale at 2~cm and 3.5~cm. Follow-up imaging with MERLIN did 
not detect this source with a 0$^{``}$.04 beam, indicating that the radio 
emission is generated in a region with a radius of $\approx 25$~au, which is 
broadly similar to the radius of the bipolar cavities inferred from models of 
near infrared data. Interpretation of this system is complicated by the 
existence of a quadrupolar outflow, i.e. two bipolar outflows along roughly 
perpendicular axes, which we originally detected through polarimetric 
imaging. We present a near infrared H$_{2}$ image in which a bow shock
in the secondary outflow is clearly seen. This complicated structure may
have been caused by a gravitational interaction between two protostars.

\twocolumn

\setlength{\parindent}{4mm}
\section{Introduction}

	For many years it has been known that some Young Stellar Objects 
(YSOs) are sources of radio emission in the cm continuum. The more deeply 
embedded YSOs, now commonly referred to as Class I and Class 0 protostars, 
exhibit the `thermal' spectrum of free-free radio emission between 2 and 
6~cm. At shorter wavelengths continuum emission from cold circumstellar dust 
becomes the dominant source of radiation (Rodriguez 1994). The energy 
source for the ionisation and free-free emission in low mass YSOs is still 
unknown. As the sensitivity of radio telescopes has improved, free-free 
emission has been detected from an increasing number of protostars, 
particularly those with CO outflows and associated Herbig-Haro flows. 
Rodriguez and Reipurth (1996) suggested that that all such systems may be 
radio sources. 

	We have conducted a sensitive survey of a small sample of Class I
YSOs in the Taurus molecular cloud which exhibit outflow activity. The aim
was to discover whether significant radio emission is indeed a universal 
feature of such systems and, by comparison with near infrared data, to shed 
light on the process which produces it. By observing at high spatial 
resolution we also hoped to probe the structure of the circumstellar matter 
on the scale of the solar system. The radio emission might trace the structure
because shock-ionisation of the outflow where it is collimated by the 
molecular envelope is a long-standing hypothesis to explain the free-free 
emission. It transpired that most Class I sources are too faint for detailed 
mapping to be achieved with present radio receiver technology but some useful 
spatial information was obtained.

	The initial survey and follow-up observations at radio and infrared
wavelengths are described in Section 2. The results are detailed in Section 3 
and in Section 4 we discuss the possible explanations for the differing radio 
fluxes of otherwise similar protostars. Our conclusions are presented in
Section 5.

\section{Observations}

\subsection{Radio Observations}

	We selected a sample of seven low mass YSOs from the nearby 
Taurus-Auriga star forming region which we have previously studied in the 
near infrared waveband. These systems all display Herbig-Haro jets or knots 
of ejected nebulosity at near infrared or optical wavelengths (Lucas \& Roche 
1997 (hereafter LR97); 1998 and references therein) and all possess CO 
outflows (Moriarty-Schieven et al. 1992). Bolometric source luminosities are
given in Table 1. We will refer to these sources by their IRAS coordinate 
designations. One of the eight sources from the near 
infrared sample, 04187+1836 (L1551-IRS5), was not observed since it had 
already been the subject of several radio investigations 
(Bieging \& Cohen 1985; Rodriguez et al. 1986).

	The initial observations were conducted with the VLA in A
configuration on 14 December 1996. Each source was 
observed in X band (3.5~cm), the most sensitive wavelength for these 
observations, with on-source integration times of 36 minutes. Three sources 
were also observed with varying integration times in U band (2~cm) in order to 
provide higher spatial resolution and information about the spectral energy 
distribution. Two of these three sources (IRAS 04016+2610 and IRAS 04361+2547) 
had previously been detected at cm wavelengths (Rodriguez et al. 1989;
Terebey, Vogel \& Myers 1992) but none of the 
others had previously been studied in this waveband. The observations
of each source were separated in time in order to provide
good $uv$ plane coverage. The flux calibrator was 3C48 and the 
phase calibrator was 0400+258, the latter being observed every 
10-15 minutes to permit accurate phase calibration. Atmospheric conditions were
stable, permitting the collection of high quality data in both X band and
U band. In A configuration the spatial 
resolution is $\approx 0.22$ arcsec (FWHM) at 3.5~cm and $\approx 0.13$ arcsec
at 2~cm. At the distance of the Taurus molecular cloud, 140~pc, 0.22 arcsec
corresponds to 30~au.

	Roughly simultaneous observations were conducted with MERLIN 
between 20 November and 17 December 1996 and follow-up observations
were made later with the VLA. 04016+2610 (L1489 IRS) and 04302+2247 
were observed with MERLIN and so was HL Tau, a previously known radio source,
at the request of the MERLIN Panel for the Allocation of Telescope Time. 
These observations were made between 20 November and 17 December 1996, with 
integration times of between 10 and 16 hours
in each case. Six elements of MERLIN were used, observing at C band (6~cm).  
The flux calibrator was 3C48 and the phase calibrator was 0400+258.
At this frequency MERLIN has a beam size of 0.04 arcsec (6~au at 140~pc), so 
it is most sensitive to very compact emission on sub-solar system scales
for these objects.

	Deeper follow-up mapping with the VLA was conducted on 29 May 1998, 
again in A configuration. 04016+2610 was observed in U band, in order to 
maximise spatial resolution at the expense of sensitivity, with an on-source 
integration time of 4 hours and 49 minutes. 04302+2247 and 04239+2436 
were observed at X band, each with an integration time of 1 hour and 48 
minutes. The flux calibrator was 3C286 and the phase calibrator was 0400+258. 
Observations of HH30, a system famous for its directly observed accretion 
disk and prominent bipolar jet, were made with the VLA in D configuration
on 17th March 1999. The integration time was 4 hours, observing at X band  
for best sensitivity. The flux calibrator was 3C48 and the phase calibrator 
was 0400+258. All the datasets were reduced and analysed with the AIPS 
software package, using standard techniques. 

\subsection{Near Infrared Observations}

	Near infrared observations were made with the 3.8-m United Kingdom 
Infrared Telescope (UKIRT). 04016+2610 was observed with UFTI, the UKIRT Fast 
Track Imager, on 5 October 1998 during its commissioning period at UKIRT. 
UFTI is a common-user near infrared camera with a 1024$^{2}$ HgCdTe HAWAII 
array, designed for high resolution 
imaging with the recently upgraded UKIRT in the 0.8-2.5~$\mu$m region. 
UFTI was constructed at Oxford University by a team led by the authors. 
Descriptions of the instrument design and of operational details are available 
on the world wide web (Roche \& Lucas 1998 and Leggett 1998). 
04016+2610 was imaged in the (1-0) S(1) line of H$_{2}$ at 2.12~$\mu$m and in 
K band, in order to probe the unusual outflow structure of the system. The 
new K filter optimised for Mauna Kea, which will be employed by the Gemini 
telescopes, was used. It is very similar to the standard Barr K filter 
(2.00-2.40~$\mu$m). The integration times were 900~s and 580~s in the H$_{2}$ 
and K filters respectively. Thin and variable cloud was present but the 
throughput was not seriously diminished. Seeing conditions were fairly good, 
with a FWHM of 0.7 arcsec seen for field stars even though the tip/tilt 
function of the UKIRT secondary mirror could not be used inside the L1489 
dark cloud, due to lack of optically bright guide stars. The image scale is 
0.091 arcsec/pixel. Median filtered flatfields were constructed from nine 
point image mosaics. The core of the source was 
close to saturation in the K band data, leading to a loss of photometric 
accuracy at the flux peak. In this paper we examine the extended outflow 
structure so this is of little concern. The observers were PFR, PWL and 
Sandy Leggett.

	04016+2610 was also observed in the L$^{\prime}$ band 
(3.5-4.1~$\mu$m) with IRCAM-3, the other near infrared camera at UKIRT, 
which has a 256$^{2}$ InSb array, which is sensitive up to 5.5$\mu$m in
the thermal infrared. The integration time was 240~s and the plate scale was
0.284 arcsec/pixel. These observations were made by Sandy Leggett on 
30 September 1998 as part of the UKIRT Service programme.

	A 2$~\mu$m spectrum of 04302+2247 was obtained with CGS4, the near 
infrared spectrograph at UKIRT, on 30 November 1999. This system has not
been included in previous spectroscopic surveys of YSOs, so the spectrum was 
obtained to investigate outflow activity via the H$_{2}$ and Brackett~$\gamma$
emission lines at 2.12 and 2.17~$\mu$m respectively. CGS4 has an InSb array
of the same type as IRCAM-3. The 40~l/mm grating was used, giving a
dispersion of 2.5~nm per pixel and spectral coverage from 1.9 to 2.5~$\mu$m. 
The slit width was 2 pixels, corresponding to 1.2 arcsec on the sky, and 
the slit position angle was 87.1 degrees, roughly parallel to the axis of the 
circumstellar envelope. The effective spectral resolution was 440 and the 
array was stepped twice at each sky position in a 1 pixel step to remove bad 
pixels. The integration time was 672~s. All of the infrared datasets were 
reduced with the IRAF software package.

\section{Results}

\subsection{The Initial Survey}
\subsubsection{VLA Data - 1st Epoch}

	Of the seven sources observed one, 04016+2610 (L1489 IRS),
was clearly detected at 3.5~cm and 2~cm. However the 2~cm emission is 
visible only when the data are tapered to the resolution of the 3.5~cm data, 
$\approx 0.22$ arcsec, indicating that this source may be resolved out on 
smaller scales. Three other sources were weakly detected at 3.5~cm at the 
4.5-5.0-$\sigma$ level: 04361+2547 (TMR-1), 04302+2247 and 04239+2436. At 
2~cm, 04361+2547 
and 04302+2247 were observed and point sources were marginally detected at the 
same positions with even lower signal to noise (see Table 1). The three weak
detections are consistent with emission from point sources but
extended emission cannot be ruled out with confidence.
The other 3 sources possess no radio counterparts to a 3-$\sigma$ limit of 
$\approx 0.10$~mJy/beam. The noise in interferometer data is non-Gaussian and 
can behave unpredictably. Consequently, it is common to require 5-$\sigma$ 
for a secure detection but a 3-$\sigma$ limit can be taken as a reliable 
non-detection when the position is known. In the case of 04302+2247, a 
combination of Digital Sky Survey and archival Hubble Space Telescope data 
allowed us to determine that the radio candidate lay within 1 arcsec of the 
faint optical counterpart, indicating an almost certain correspondence. For 
04239+2436, the published infrared positions were only accurate to 3 
arcseconds so confirmation had to await the second, deeper set of VLA data.

	The maps of 04016+2610 are displayed in Figure 1(a-b), and a 
near infrared comparison image in polarized intensity (surface brightness 
multiplied by degree of polarization) is reproduced from LR97
in Figure 2. At 3.5~cm we see a point source with extension to the northwest 
and at 2~cm the structure is similar but the extended emission is only 
marginally detected in approximately the same direction. The spatial scale of 
the extension is $\approx 0.4$ arcsec (50~au) and we see that it is not 
aligned with the outflow cavities visible in the infrared image. In LR97 we 
interpreted the near infrared structure as a `quadrupolar outflow', i.e. two 
perpendicular bipolar flows, perhaps indicative of an unresolved binary system.
The approximate directions of the primary and secondary outflows (so named
for their relative prominence) are marked in Figure 2.

	Two point spectral indices can be very roughly estimated for the 3
sources for which we have both X and U band data, using the data in Table 1.
In each case the data imply a small positive index: $0 < \alpha < 2$
where the flux density S$_{\nu}$ at frequency $\nu$ is described by 
S$_{\nu} \propto \nu^{\alpha}$. This is consistent with free-free emission 
from partially ionised gas, rather than the continuum emission from cold
dust. Gyro-synchrotron emission cannot be ruled out entirely but no circular 
polarization is detected in 04016+2610 above a 3-$\sigma$ limit of 32\% 
at 3.5~cm.

\subsubsection{MERLIN Data}

	None of the three sources which were observed with the 0.04 arcsec
MERLIN beam were detected at 6~cm (see Table 1). 04016+2610 has a flux
of 0.5 mJy at 6~cm, measured by Rodriguez et al.(1989) at low spatial
resolution, but it was not detected to a 3-$\sigma$
limit of 0.26~mJy/beam. The non-detection of 04016+2610 is therefore
significant and demonstrates that the radio emission arises on spatial scales 
of tens of astronomical units, comparable to the size of the solar system. 
This is consistent with the VLA 2~cm data, which also appear to be on the
verge of resolving out the emission, as described above. Since the VLA
and MERLIN observations were obtained within a few days of each other it
is unlikely that source variability could account for the MERLIN 
non-detection. HL Tau has a 6~cm flux of only $\approx 0.22$~mJy (Wilner et 
al. 1996) so the MERLIN data do not provide useful information on this object.
Similarly, the faintness of the 3.6~cm emission from 04302+2247 
is consistent with the non-detection by MERLIN, assuming a thermal spectrum.

\subsection{Further Observations}
\subsubsection{VLA Data}
	
	The 1998 observations of 04302+2247 and 04239+2436 confirm the
detections of these sources (see Table 1) at the same location as the 1st
epoch candidates. The fluxes measured at the 2 epochs are the same within the 
1-$\sigma$ errors and both systems appear to be unresolved on a scale of 
30~au. 04302+2247 is the lowest luminosity YSO from which radio emission has 
so far been detected, with L$\mathrm{_{bol} \approx 0.3}$~L$_{\odot}$, and 
04239+2436 is only slightly more luminous with L$\mathrm{_{bol} \approx 
1.2}$~L$_{\odot}$ (Tamura et al. 1991). The low luminosities make it most 
unlikely that photoionisation is responsible for ionising the outflow,
(see Section 4.2.2).

	In the 1998 data the morphology of 04016+2610 appears to have 
changed: only a point source is detected, with no sign of the extended 
emission desite the greater sensitivity of the observations. The source 
remains visible at the full resolution of the VLA at 2~cm, which is 
0$^{``}$.136 $\times$ 0$^{``}$.113 in a ``robust zero'' map (which adds 
some weight to the longer baselines with only a minimal reduction in signal 
to noise, relative to a naturally weighted map). The image profile is not 
perfectly symmetric but the JMFIT beam-fitting task in 
AIPS indicates that the data are consistent with a point source. It is 
probable that this is a real change in source morphology, related to the 
outflow activity of the protostar but the peak flux density does not appear 
to have changed between the two epochs.

	The deep integration at the location of HH30 in 1999 failed to 
detect this system to a 3-$\sigma$ limit of 0.03~mJy at 3.5~cm. This system
was also undetected to a limit of $\approx 0.1$~mJy in 3.5~cm observations
of the HL Tau/HH30 region made in 1992 by Luis Rodriguez (private 
communication).

\subsection{Infrared Data}
\subsubsection{04016+2610}

	The K band and H$_{2}$ images of 04016+2610 are displayed in 
Figure 3(a,b). The most striking feature in H$_{2}$ is a well defined, 
V-shaped, emission nebulosity to the south of the core, which we interpret as 
a bow-shock in the secondary outflow of the system. In the continuum image 
this bow shock is mostly obscured by the surrounding reflection nebulosity.
The nebulosity also extends several hundred au to the north and a resolved
continuum source (IRS2) is visible close to the core in this direction, 
coincident with the feature seen in polarized intensity in Figure 2. 
These features support the inference of a secondary bipolar outflow along the 
direction indicated in Figure 2. It is oriented roughly perpendicular to the 
primary east-west outflow but is not clearly visible in existing high resolution 
imaging (non-polarimetric) data (LR97 and Padgett et al. 1999). 
Although IRS2 appears rounded in our relatively low resolution data, the high 
resolution 3-colour NICMOS data of Padgett et al. reveal that IRS2 has a 
triangular shape which is consistent with a conical cavity in the nebula that is
illuminated by a source at or close to the central flux peak. 
The quadrupolar outflow structure is also indicated in the CS (J=2-1)
map of Ohashi et al.(1996), although the velocity distribution of the
low density gas depicted there is more consistent with \it{infalling}
\rm material. 

	The H$_{2}$ and K images were normalised such that the same flux was 
received in both filters from a nearby field star in the images before 
subtracting to produce the image in Figure 3(c). Since the field star lies
within or behind the L1489 dark cloud, the effect of the extinction
by the cloud is cancelled to some degree. Hence, the solid (positive) contours 
represent nebulosity which appears bluer than the field star in the
(H$_{2}$-K) colour, due either to H$_{2}$ emission or a bluer K band continuum, 
while the dotted (negative) contours represent redder 
radiation. The bow shock is the only feature in Fig. 3(c) which we
attribute to H$_{2}$ emission. The rest of the structure in Fig.3(c) is
real, except at the flux peak (marked with an asterisk), which appears
positive owing to saturation at K band. The blue colour of the
southeastern nebulosity is consistent with Rayleigh-like scattering in
the K band (F$_{\lambda} \propto \lambda^{-4}$), which was indicated by
the very high ($>70\%$) degrees of linear polarisation measured in this 
system (LR97). The red nebulosity near the flux peak corresponds to
the densest material where the scattered light is strongly reddened by 
extinction. Our interpretation of the residuals as variations in colour 
across the K band (2.0-2.4~$\mu$m) is consistent with the Padgett et al. 
data, which also shows that the reddest nebulosity lies to the northwest, while 
the southeastern material is bluer. If the system is a binary then it could 
either be unresolved by NICMOS (separation $\ltsimeq 0^{``}.1=14$~au, 
depending on the relative brightness) or an unseen component may lie 
obscured behind the very red, dense material to the northwest of the central 
flux peak. This latter possibility is suggested by the shape of the putative 
conical cavity, IRS2, which appears to focus on a point at least 1 arcsec 
northwest of the primary flux peak.

	The L$^{\prime}$ data in Figure 4 show a cometary nebula similar in 
appearance to the K band continuum data, with no sign of the secondary 
outflow. It appears that sub-arcsecond millimetre continuum data will be needed
to discover the distribution of matter in this unusual system.

\subsubsection{04302+2247}

	The 2~micron spectrum of 04302+2247 (Figure 5) rises toward longer
wavelengths and shows emission lines of Br$\gamma$, Br$\delta$ and the
H$_{2}$ (1-0)S(1) transition. None of the other H$_{2}$ lines are clearly
detected. The equivalent widths are $\mathrm{EW_{\lambda}(Br\gamma)=14.5 \pm 
2.1 \AA}$ and $\mathrm{EW_{\lambda}(H_{2})=5.2\pm 1.0\AA}$ respectively. 
Br$\delta$ is blended with a telluric absorption feature and is hard to measure. 
The Br$\gamma$ equivalent width is exceptionally large for a Class I protostar, 
but the H$_{2}$ line strength line is only slightly above average 
(see Greene \& Lada 1996a). The Br$\gamma$ equivalent width is similar to that 
of 04239+2436 (Greene \& Lada 1996b) which is notable for its extreme near 
infrared spectrum. However, 04302+2247 differs from 04239+2436 in that it does 
not show the CO (v=2-0) transitions in emission, only marginally detected
features in absorption.
The core of the system is obscured at near infrared wavelengths and the 
Br$\gamma$ emission is thought to originate in the hot, dense conditions 
close to the star, so we presume that this line is observed in scattered 
light. The strong Br$\gamma$ but weaker H$_{2}$ implies that the wind of 
04302+2247 has an unusually high temperature or ionisation fraction.
 
\subsubsection{04239+2436}

	In an earlier paper (Lucas \& Roche 1998) we presented broad band 
imaging polarimetry of this system and noted that a possible jet-like feature 
was marginally detected along the symmetry axis of the circumstellar envelope
but was mostly obscured by the nebulosity. High resolution near infrared 
images of this system were obtained with NICMOS on the Hubble Space Telescope 
(HST) in 1998 (PI Reipurth, data in preparation). The 1.6$\mu$m broad band 
image in the HST archive (not shown) clearly resolves a well collimated 
knotted jet along the symmetry axis. This prominent outflow is clearly 
related to the extreme outflow activity implied by the near infrared 
spectrum of Greene \& Lada. The detection of radio emission in two systems 
with exceptionally strong HI emission lines, 04239+2436 and 04302+2247, is 
interesting. However, the radio sources 04016+2610 and 04361+2547 have 
unremarkable HI emission.

\section{Discussion and Interpretation}

\subsection{04016+2610 and Cavity Radii}

	The observation that radio emission in 04016+2610 arises on a scale 
of 50~au, or 25~au radius, is important because it sets a lower limit 
on the cavity radius in the core, where the two bipolar cavities presumably 
merge into a single cleared space. This important parameter was inferred from 
near infrared Monte Carlo modelling of YSOs as being about 25~au (eg. LR97) 
but has never been well constrained. Admittedly 04016+2610 is far from an 
ideal test case, being a very complicated system and probably a binary, but 
this is also the spatial scale of the gyrosynchrotron emission observed 
around T Tau South by Ray et al.(1997) using MERLIN. Hence we can have a 
little more confidence in inferences drawn from near infrared simulations.

	We speculate that the origin of the strange quadru-polar outflow 
geometry in 04016+2610 lies in a gravitational interaction between stars
which formed in separate protostellar cores and later underwent a close
encounter. Hydrodynamical simulations of such scenarios have been performed 
by Boffin et al.(1998) for star-disc encounters and Watkins et al.(1998a, 
1998b) for disc-disc encounters. These simulations show that a number of 
bodies may be produced in such encounters, and their rotation axes need not be 
parallel.  

\subsection{The Criterion for Detectable Radio Emission}

	The reason why some YSOs emit detectable centimetre continuum 
flux and others do not has been puzzle since the 1980s. Gibb (1999)
recently reviewed the problem using a large dataset and showed that
radio emission is relatively common among young (Class I and Class 0) sources
and among chromospherically active Class III sources, which are presumed
to emit gyrosynchrotron radiation, but rare in Class II systems. We are 
concerned here with the younger group of radio emitters, all of which drive 
molecular outflows. Rodriguez (1994) stated that 40/180 of the then known 
molecular outflow sources were radio emitters, and suggested that this 22\% 
fraction may increase as more sensitive surveys are carried out. Gibb suggested 
that the polar orientation of the system might be an important factor, owing to 
self-absorption	in optically thick systems viewed close to the outflow axis. 
This probably does have some effect but since the great majority of YSOs are 
not viewed close to pole-on there must be an additional explanation for the 
non-detection of a large fraction of Class I sources.

\subsubsection{Variability}

	Reviewing the results of our initial radio survey and the
confirmatory detections of 04302+2247 and 04239+2436 in the 2nd epoch, we 
note that the outflows of the 4/7 detected sources all display good evidence 
for an ongoing or recent ejection event. In the infrared, 04016+2610 has a 
marginally resolved secondary continuum source close to the core - probably 
a knot of gas and dust in the secondary outflow. 04239+2436 has
extremely active near 
infrared line emission and a powerful Herbig-Haro jet. 04302+2247 has a 
quadrupolar structure in the infrared, which has been attributed to 
outflowing material within 200 au of the core, either along the outflow axis 
(Lucas \& Roche 1997) or flowing along the cavity walls (Padgett et al. 1999). 
Finally, 04361+2547 (TMR-1) is a resolved binary with an asymmetric core 
structure and a possible low mass companion which appears to be
connected to the system by a long channel of reflected light (Terebey et
al. 1998,2000). In this
case outflow actvity is hard to separate from the confused nebular strcture.
Admittedly, the undetected radio sources also display infrared outflow 
activity but in the form of quite discrete Herbig-Haro knots rather than a 
jet. Hence it is possible that the source of the outflow was quiescent at the 
time of observation.

	The apparent radio variability of 04016+2610 suggested the 
following scenario: a Class I protostar ejects a roughly spherical knot of 
partially ionised gas at intervals of a few years, which expands
and becomes more rarefied until it is optically thin to free-free emission.
As shown by Martin (1996) the radio flux from such an isothermal stellar wind 
or knot is given by:

(1) $S_{\nu}(T)=\frac{2k_{B}T\nu^{2}}{c^{2}d^{2}}\int_{A} (1-e^{-\tau(\nu,T)})dA^{\prime}$\\

(2) $\tau=\int \kappa(\nu,T) n^{2}_{e} ds$

assuming that the Planck function is appropriate in the optically thick
limit. Hence, the flux from an expanding knot will rapidly decline when the
the optically thin threshold is passed. Unfortunately, the electron number 
density, $n_{e}$, is very poorly constrained for knots near the point of 
origin: for continuous wind models Martin found that models with an electron 
density at 1~au from the star in the range $10^{7}$ to $10^{12}$~cm$^{-3}$ 
can match observed fluxes and spectral indices. Moreover, radio variability 
data on timescales of a few years is very sparse. 

	To test this idea we undertook the observation of the Class II system
HH30. It has a very prominent, well collimated jet with knots observed 
within a few dozen au of the star (Burrows et al. 1996), apparently ejected at 
intervals of order 1 year and the jet can be observed to within 30~au 
of the source, making it ideal for comparative radio/optical studies.
(We assume here that the knots are discrete ejections rather than shocks in a 
steady flow). However, no radio source was detected to the stringent limit of 
0.03~mJy/beam at 3.5~cm in 1999, nor in 1992 above 0.1~mJy/beam. A 
similar but younger system (Class I or Class 0) is HH212 (Zinnecker, 
McCaughrean \& Rayner 1998). HH212 also has a well collimated and knotted jet 
in which the perfect 
bipolar symmetry of the knots indicates that they do represent discrete 
ejections of matter. HH212 might be expected to have a higher rate of outflow 
and a stronger radio flux given its evolutionary status. However, Zinnecker
et al. also detected no radio source to a 4-$\sigma$ limit of 0.06~mJy/beam. 
These non-detections could simply be due to observation in a quiescent period 
of outflow activity, especially in the case of HH212, where the knots are 
ejected at 5 year intervals. However, we must conclude that the radio 
variability hypothesis is given no support by these limited data, and large 
amplitude radio variability has not, to our knowledge, yet been observed in 
these systems.

\subsubsection{Ionisation}

	Another possibility suggested by eq.2 is that the number density
of electrons is low in radio quiet systems due to a low ionisation fraction.
Reynolds (1986) calculated that $S_{\nu} \propto n_{e}^{1.67}$ for an
isothermal collimated wind. This is a slightly weaker dependence than the 
$n_{e}^{2}$ relation for an optically thin wind but an order of magnitude 
reduction in radio flux would be caused by less than
a factor of 4 difference in the ionisation fraction. The other important 
variable, temperature, has a relatively weak effect on the source
function. As mentioned in Section 1, the origin of the ionisation is unknown. 
Photoionisation by the photospheric UV radiation is insignificant for
cool low mass YSOs. Boundary layer emission where matter accretes 
on to the stellar surface does not seem likely, given that modelling of the 
UV continuum in classical T Tauri stars by Gullbring et al.(1998) indicates 
temperatures of only about 10000~K. 
Shock ionisation of the wind remains a serious possibility. Delamarter,
Frank \& Hartmann (2000) have conducted numerical simulations of the 
interaction of a spherical wind and an equatorially condensed circumstellar 
envelope. Their results show that physical collimation of such a wind by
the envelope does produce cavity structures consistent with near infrared
and millimetre imaging data. Bacciotti
\& Eisloffel (1999) have directly and precisely measured the ionisation of 
the knots in several stellar jets, under the assumption that the ionisation 
states of nitrogen and oxygen are maintained by charge transfer with
hydrogen. They have shown that the ionisation fraction, $x_{e}=n_{e}/n_{H}$, 
typically lies in the range $x_{e}=5$ to $35\%$ but can be as low as 1.5\% 
(in parts of the HH34 jet). Denser and less excited jets tend toward the 
smaller values. In general, for a given section of the jet the ionisation 
fraction appears to decline 
slowly with increasing distance from the source but the fraction in separate 
sections is uncorrelated. Thus, a newly ejected knot can have a low level of 
ionisation, as in the case of HH24E for example. Clearly, a negligible
ionisation fraction is not consistent with the prominent H$\alpha$ and
[SII] emission which has been observed with HST.

	The best available spectroscopic data on the HH30 jet is given by
Mundt et al.(1990), which is relatively low resolution ground-based data,
but includes the necessary [OI], [NII], [SII] and H$\alpha$ lines. 
We attempted to apply the Bacciotti method to the printed data and a low 
ionisation fraction appeared plausible, but the problem has very recently been
fully addressed by the pioneers of this technique, see Bacciotti, 
Eisloffel \& Ray (1999).
They combined the Mundt et al. spectra with HST emission line images from 1995
to probe conditions with very fine resolution, with the 
caveat that possible temporal changes in electron density and excitation may 
cause errors. They find $x_{e}=6.5\%$ at the base of the jet and that 
$x_{e}$ actually rises slowly to 14\% at 300~au from the source. The low
ionisation fraction at source supports our hypothesis that this may
be the reason why the system is radio quiet. We must caution that they have
measured $x_{e}$ in relatively low density conditions 
($n_{e} < 10^{4}$~cm$^{-3}$), compared to those which must prevail on
small spatial scales at the source of the jet.
The hydrogen gas density in the jet measured by Bacciotti et al. is unusually
high compared to other systems, which 
indicates that the very low radio flux cannot be attributed simply to 
insufficient material. In their analysis they also argue that the observed
bright knots are not regions of high density but merely to local shocks. This
is a controversial point but if true it would mean that gas density is
sufficient for radio emission at all times.

	Is the ionisation fraction therefore the discriminating factor for
radio emission from Class I sources? If we take $x_{e}=15\%$ as the average
value and 6.5\% in HH30 then the system might be a factor of 5 fainter 
than average, assuming $S_{\nu} \propto x_{e}^{2}$. The actual dependence of 
flux density on $x_{e}$ is a function of the wind geometry in the core, as 
described by Reynolds (1986), and may well be weaker. 3/4 of the radio 
protostars detected in our Taurus sample have fluxes of only 0.10-0.15 mJy, so
low ionisation may be just sufficient to account for a flux density below our
upper limit of 0.03~mJy/beam, particularly if the ionisation fraction is even 
lower than normal in the dense radio emitting region. If so, then why is 
the ionisation fraction low in HH30 ? Both HH30 and HH212 have very well 
collimated jets and the observed disk of HH30 is confined to equatorial 
latitudes. Thus the jet might avoid shock-ionisation by interaction with the 
surrounding envelope while poorly collimated outflows such as that of 
04302+2247 do not. 
It is hard to assess the collimation of most of the outflows in our sample 
since in most cases the infrared nebulosity prevents clear delineation of the 
outflow. 04239+2436 has a well collimated jet in the NICMOS data and is a 
radio source but poorly collimated optical [SII] emission was observed by 
Gomez et al.(1997) to one side of the system's axis of symmetry. Some well 
collimated jets (eg. HH34) are radio bright but their envelope structure is 
largely obscured from view. Direct imaging of jet/envelope interactions
will have to be undertaken using adaptive optics on 8-m class 
telescopes.

\section{Conclusions}

	Observation of a small sample of low mass Class I protostars
indicates that radio emission at above the level of 0.1~mJy/beam  is far 
from ubiquitous. We suggest that the radio quietness in sources with
active Herbig-Haro flows might be accounted for by a relatively low 
ionisation fraction, at the level of a few percent. The ionisation
fraction could plausibly be related to the collimation of the outflow,
particularly if shock-ionisation by interaction with the circumstellar 
envelope plays a role. In addition, radio variability due to unsteady 
outflows may also occur and this should be investigated by 
monitoring known radio emitters.

	In 04016+2610 (L1489 IRS) a well defined bow shock is observed 
in 2.12~$\mu$m H$_{2}$ radiation along the line of the secondary bipolar 
cavity revealed in a polarised intensity image. This provides further 
evidence for a quadrupolar flow. We suggest that the system is the product of 
a gravitational interaction between two young solar systems, which produced 
a binary star whose components drive outflows along different axes of 
rotation. The system may be an unresolved binary (separation $\ltsimeq 
0^{``}.1$ or 14 au), or the secondary component may be obscured by dense 
material in the core of the system. 

\section{Acknowledgements}

	We are grateful to the VLA Time Allocation Committee for awarding
us substantial amounts of time for this project. We especially wish to 
thank Mike Rupen of the VLA Array Operations Centre, Socorro for his
help in calibrating and reducing the data, and also his colleagues for 
their hospitality. We also thank the MERLIN Time Allocation Committee
giving us time to set some upper limits and Simon Garrington for his help
in reducing the data. We also wish to thank everyone at Oxford, Edinburgh
Cambridge and the Joint Astronomy Centre who participated in the design,
construction and commissioning of UFTI at UKIRT. Special thanks go to Tony 
Handford, Keith Nobbs and Phil Evans of the Nuclear and Astrophysics Department 
workshop in Oxford, for transforming the designs into real metal.
UKIRT and the Joint Astronomy Centre are operated on behalf of the UK
Particle Physics and Astronomy Research Council.

\vspace{1cm}

\large{\textbf{References}}\textmd{}\normalsize
\setlength {\parskip} {2mm}
\setlength {\parindent} {0mm}

Bacciotti F., \& Eisloffel J. 1999, A\&A 342,717 \\
Bacciotti F., Eisloffel J., \& Ray T.P. 1999, A\&A 350,917 \\
Bieging J.H., \& Cohen M. 1985, ApJ 289, L5\\ 
Boffin H.M.J., Watkins S.J., Bhattal A.S., Francis N., \& Whitworth A.P. 1998, 
MNRAS 300,1189\\
Burrows C.J., Stapelfeldt K.R., Watson A.M., Krist J.E., Ballester G.E.,
Clarke J.T., Crisp D., Gallagher J.S. III, Griffiths R.E., Hester J.J., 
Hoessel J.G., Holtzmann J.A., Mould J.R., Scowen P.A., Trauger J.T., 
Westphal J.A., 1996, ApJ, 473,437\\
Delamarter G., Frank A., \& Hartmann L. 2000, ApJ 530,923\\
Gibb A. 1999, MNRAS 304,1\\
Gomez M., Whitney B.A., \& Kenyon S.J. 1997, AJ 114,1138\\
Greene T.P., \& Lada C.J. 1996a, AJ 112,2184\\
Greene T.P., \& Lada C.J. 1996b, ApJ 461,345\\
Gullbring E., Hartmann L., Briceno C., \& Calvet N. 1998, ApJ 492,323\\
Leggett S. 1998, \\
www.jach.hawaii.edu/ukirt.new/instruments/ufti/ufti.html \\
Lucas P.W., \& Roche 1997, MNRAS 286,895\\
Lucas P.W., \& Roche 1998, MNRAS 299,699\\
Martin 1996, ApJ 473,1051\\
Moriarty-Schieven G.H., Wannier P.G., Tamura M., \& Keene J. 1992, ApJ 400,260\\ 
Mundt R., Buehrke T., Solf J., Ray T., \& Raga A.C. 1990, A\&A 232,37\\
Ohashi N., Hayashi M., Kawabe R., \& Ishiguro M. 1996, ApJ 466,317\\
Padgett D.L., Brandner W., Stapelfeldt K.R., Strom S.E., Terebey S., \& 
Koerner D. 1999, AJ 117,1490\\
Ray T.P., Muxlow T.W.B., Axon D.J., Brown A., Corcoran D., Dyson J., \& 
Mundt R. 1997. Nature 385,30\\
Reynolds S.P. 1986, ApJ 304,713\\ 
Roche P.F., \& Lucas P.W. 1998, \\
www-astro.physics.ox.ac.uk/~pwl/camera.html\\
Rodriguez, L.F., Canto J., Torrelles J.M., \& Ho P.T.P. 1986, ApJ 301,L25\\
Rodriguez, L.F.; Myers, P.C.; Cruz-Gonzalez, I.; Terebey, S. 1989. 
ApJ 347,461\\
Rodriguez L.F. 1994, RMxAA 29,69\\
Rodriguez L.F., \& Reipurth B. 1996, RMxAA 32,27\\
Tamura M., Gatley I., Waller W., \& Werner M.W. 1991, ApJ 374,L25\\
Terebey S,, Vogel S.N., \& Myers P.C. 1992. ApJ 390,181\\
Terebey S., van Buren D., Padgett D.L., Hancock T., \& Brundage M. 1998. 
ApJ 507, L71\\
Terebey S., van Buren D., Matthews K., Padgett D.L. 2000. AJ 119.2341\\
Watkins S.J., Bhattal A.S., Boffin H.M.J., Francis N., \& Whitworth A.P. 
1998, MNRAS 300,1205\\ 
Watkins S.J., Bhattal A.S., Boffin H.M.J., Francis N., \& Whitworth A.P. 
1998, MNRAS 300,1214\\ 
Wilner D.J., Ho P.T.P.,, \& Rodriguez L.F. 1996, ApJ 470,L117 \\
Zinnecker H., McCaughrean M.J., \& Rayner J.T. 1998, Nature 394,862

\pagebreak

\onecolumn
\begin{table}
\caption{\textbf{Results of the Survey} \textmd{} }

\small
\begin{tabular}{lccclll}
Source & R.A. (1950)$^{a}$ & Dec (1950) & L$\mathrm{_{bol}^{b}}$ & 3.5~cm Flux$^{c,d}$  & 2~cm Flux$^{c}$ & 6cm Flux$^{d}$\\
 &  &  &    &  (mJy) & (mJy) & (mJy) \\
 
04016+2610 (L1489) & 04 01 40.55 & +26 10 47.7 & 3.7 & $0.343\pm0.033$ & $0.55\pm0.11$,$0.52\pm0.04$ & $<0.26$\\ 
04361+2547 (TMR-1) & 04 36 09.87 & +25 47 28.7 & 2.8 & $0.14\pm0.032$ &
$\sim0.30$ ? & -  \\
04302+2247 & 04 30 16.66 & +22 47 04.4 & 0.33 & $0.16\pm0.033$,$0.13\pm0.02$ & $\sim0.33$ ? & $<0.32$ \\
04239+2436 & 04 23 54.42 & +24 36 53.3 & 1.2 & $0.14\pm0.033$,$0.10\pm0.02$ & - 
& - \\
04325+2402 & - & - & 0.70 & $<0.10$ & - & - \\
04248+2612 & - & - & 0.35 & $<0.10$ & - & - \\
04365+2535 & - & - & 2.2  & $<0.10$ & - & - \\
HH30 & - & - & ?  & $<0.033$ & - & - \\
HL Tau & - & - & - & - & - & $<0.32$\\
\end{tabular}

Notes:\\ 
a: Coordinates given are the locations of the radio sources.\\
b: Bolometric luminosities are from Tamura et al.(1991)\\
c: For 04016+2610 the fluxes given are flux densities/beam for the unresolved
component. The extended flux is not well measured. Where two fluxes are given 
these are the first and second epoch measurements respectively.\\
d: upper limits are 3-$\sigma$, in mJy/beam. 
\end{table}

\vspace{25cm}

\pagebreak
\normalsize
\onecolumn

Figure 1: Radio images of 04016+2610 (L1489) obtained with the VLA.
(a) at 3.5~cm, (b) at 2~cm. The image in (b) has been tapered to the
same resolution as (a) since the source vanishes in an untapered image. 
Beam sizes are marked. Contour levels in (a) are -0.079, -0.056, 0.056, 
0.079, 0.112, 0.158, 0.224, 0.317 mJy and the 1-$\sigma$ noise level is
0.033 mJy. Contour levels in (b) are -0.33, -0.22, 0.22, 0.275, 0.33, 
0.385, 0.44, 0.495, 0.55 mJy and the 1-$\sigma$ noise level is
0.11 mJy.

Figure 2: Near infrared image of 04016+2610 in polarised flux at 2.2~$\mu$m;
reproduced from LR97. The central source is at the location of the flux
peak in Figure 3(a), and is thought to be the protostar 
(or protostars if the system is an unresolved binary). Dense regions are 
suppressed because multiple scattering reduces the degree of polarisation.
The dashed lines mark the approximate directions of the primary and secondary 
bipolar cavities as indicated by the contours. Contour levels are 0.0049, 
0.0081, 0.013, 0.019, 0.032, 0.049, 0.065, 0.097, 0.13, 0.16, 0.20, 0.24, 
0.32, 0.41, 0.49, 0.65, 0.81, normalised to unity at the peak.

Figure 3: Near infrared images of 04016+2510. (a) K band. (b) H$_{2}$ 
emission at 2.12~$\mu$m. (c) magnified view of the difference, H$_{2}$ - K,
which has been smoothed slightly with a Gaussian ($\sigma=1$ pixel). 
Contours are spaced at intervals of $\sqrt 2$, normalised to the peak. 
A bow shock is visible in the direction of the secondary outflow in (b)
but is mostly obscured by the reflection nebula in (a). The peak in (a)
is saturated and a diffraction spike produces a horizontal ridge on the
contours on either side. The structure in (c) is real, except at the location 
of the saturated flux peak (marked by an asterisk). The solid (positive) 
contours to the southeast of the peak represent regions where the scattered 
light at K band is very blue (effective wavelength $\lambda < 2.12~\mu$m), 
while the dotted (negative) contours are regions of high density where the 
scattered light has been reddened by extinction. 

Figure 4: Near infrared image of 04026+2610 at L$^{\prime}$ (3.8~$\mu$m). Each 
contour represents a step in surface brightness of a factor of 2. At this
wavelength, the nebula depicts scattered light from dense material in the 
core of the system.

Figure 5: Near infrared spectrum of 04302+2247. The red continuum is
typical of an embedded system. Br$\gamma$ and Br$\delta$ are very prominent,
but H$_{2}$ is relatively weak.

\pagebreak

\begin{figure*}
\begin{center}
\begin{picture}(200,480)

\put(0,0){\includegraphics{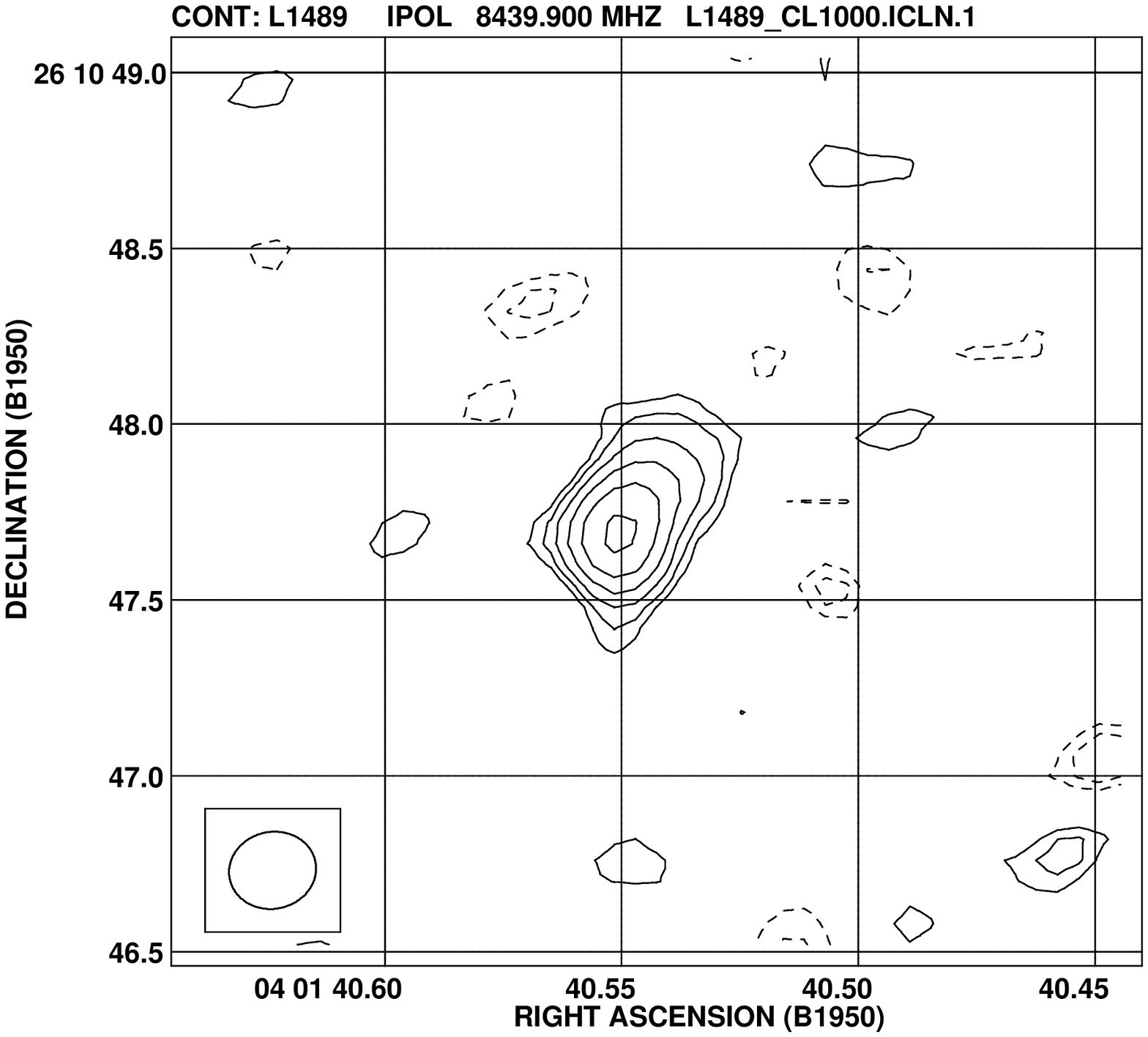}}

\put(0,0){\includegraphics{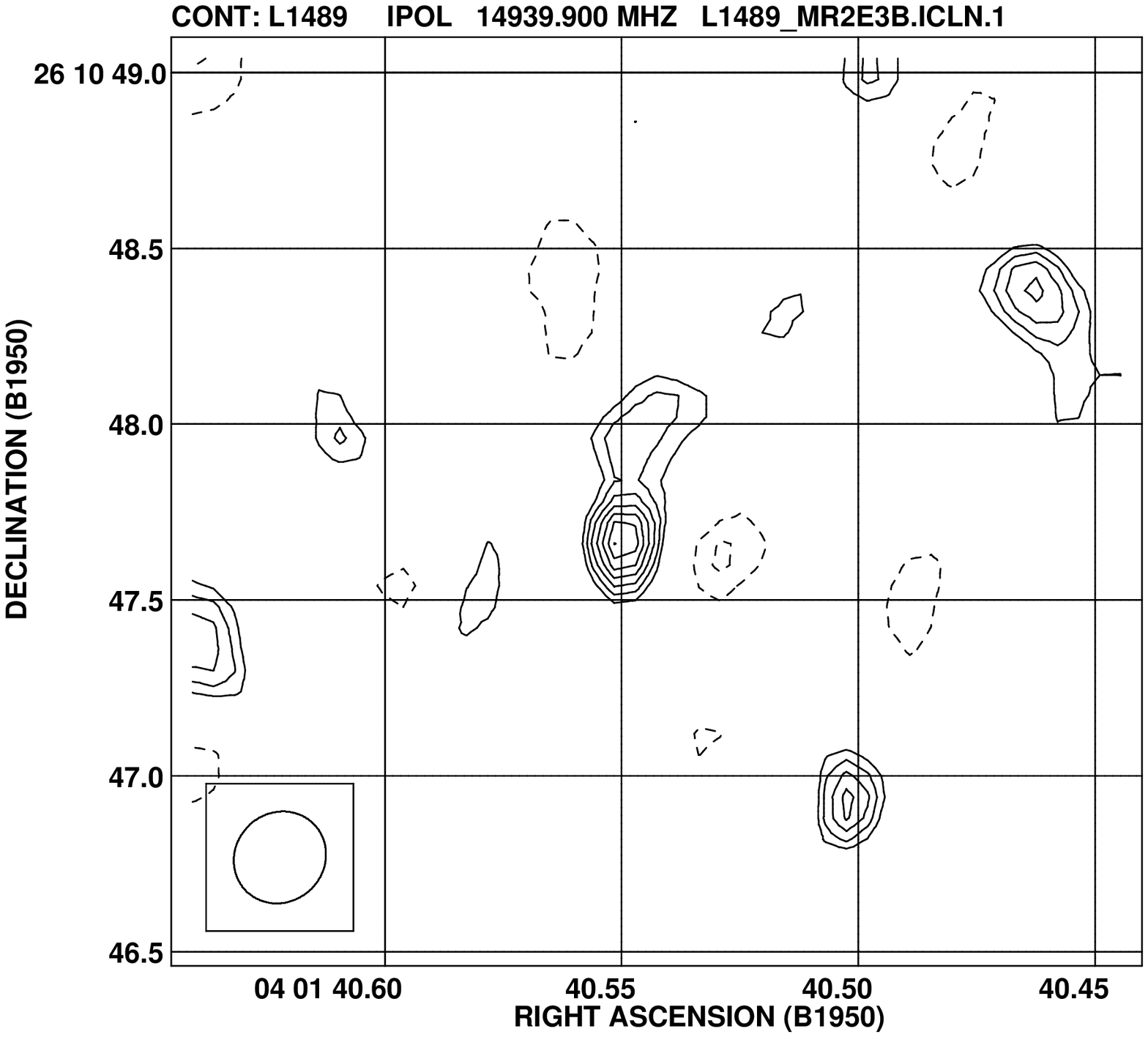}}
\put(-100,165){\textbf{(a)}}
\put(160,165){\textbf{(b)}}
\put(60,265){Figure 1}

\end{picture} 
\end{center}
\end{figure*}

\pagebreak

\begin{figure*}
\begin{center}
\begin{picture}(200,300)

\put(0,0){\includegraphics{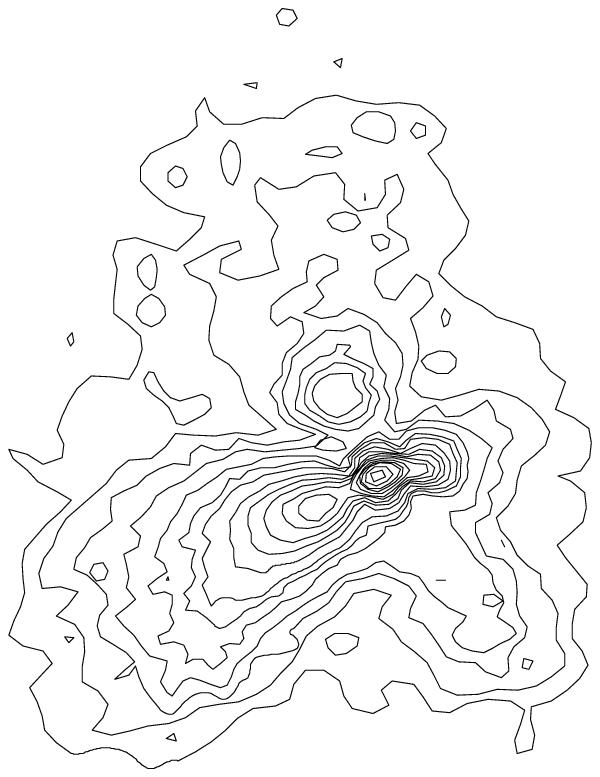}}
\put(-60,-20){\framebox(280,230)}
\put(-38,188){\textbf{04016+2610}}
\put(-36,168){\textbf{Polarized}}
\put(-36,156){\textbf{Intensity}}
\put(-30,140){\textbf{2.2~$\mu$m}}
\thicklines
\put(145,190){\line(1,0){36.8}}
\put(160,175){5''}
\thinlines
\put(149,0){\includegraphics{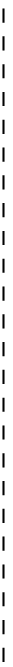}}
\put(-10,36){\includegraphics{x1.eps}}
\put(-30,22){Primary}
\put(150,-10){Secondary}

\put(190,50){\vector(0,1){40}}
\put(200,70){\line(-1,0){20}}
\put(185,95){N}
\put(165,65){E}
\put(50,250){Figure 2}

\end{picture} 
\end{center}
\end{figure*}

\pagebreak

\begin{figure*}
\begin{center}
\begin{picture}(200,480)

\put(0,0){\includegraphics{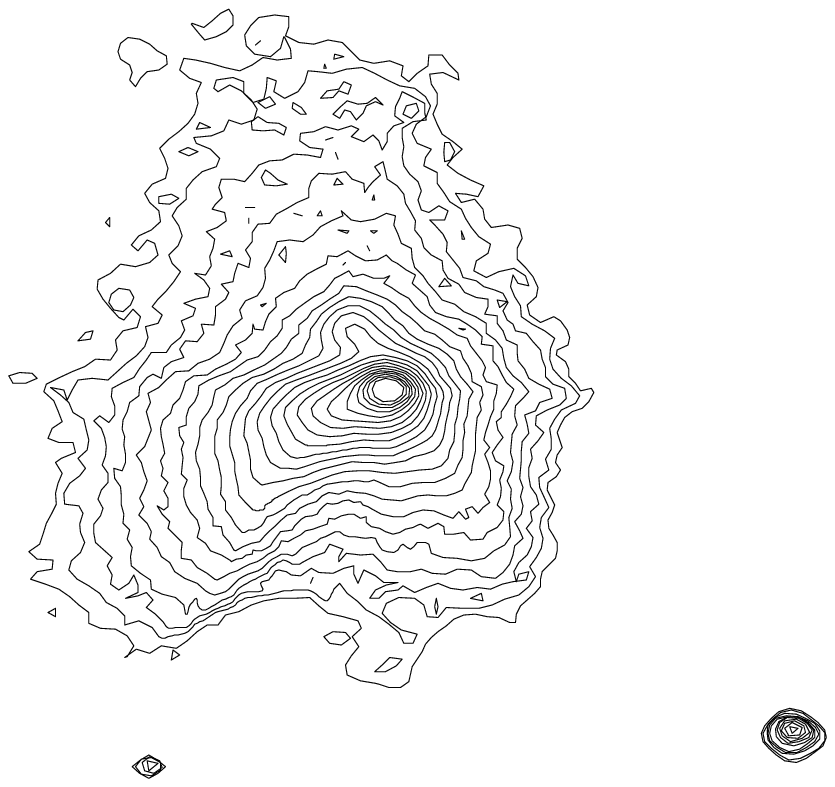}}
\put(-10,285){\framebox(300,200)}
\put(216,455){\textbf{K Band}}
\put(215,443){\textbf{(2.2~$\mu$m)}}
\thicklines
\put(225,420){\line(1,0){28.3}}
\put(235,405){5''}
\put(240,320){\vector(0,1){40}}
\put(250,340){\line(-1,0){20}}
\put(235,365){N}
\put(215,335){E}

\put(2,440){\textbf{04016+2610}}
\put(20,430){\textbf{(a)}}
\put(2,230){\textbf{04016+2610}}
\put(20,220){\textbf{(b)}}
\put(14,30){\textbf{04016+2610}}
\put(32,20){\textbf{(c)}}

\thinlines
\put(130,180){\line(2,1){40}}
\put(175,200){IRS2}
\put(75,500){Figure 3}

\put(0,0){\includegraphics{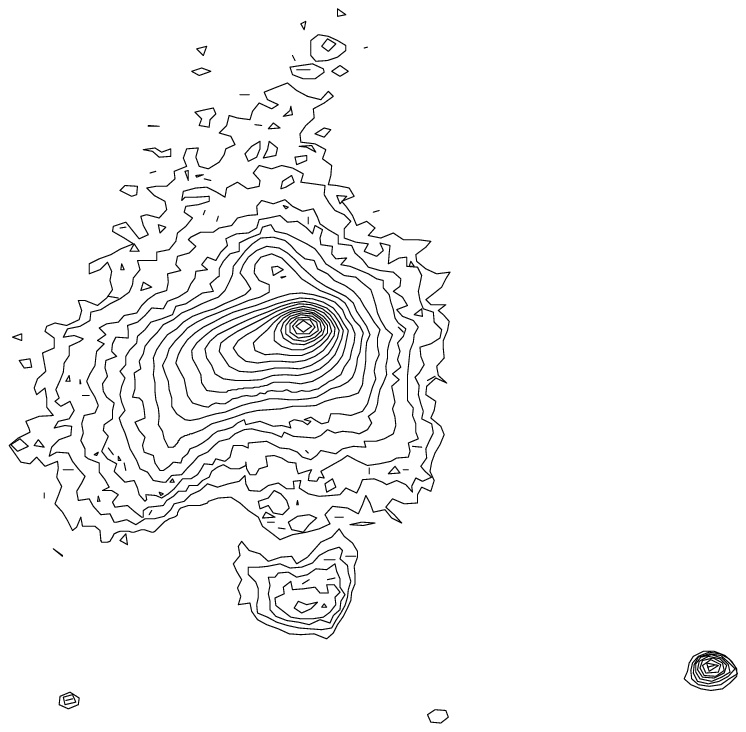}}
\put(-10,75){\framebox(300,200)}
\put(230,245){\textbf{H$_{2}$}}
\put(215,233){\textbf{(2.12~$\mu$m)}}
\thicklines
\put(225,200){\line(1,0){28.3}}
\put(235,185){5''}
\thinlines

\put(0,0){\includegraphics{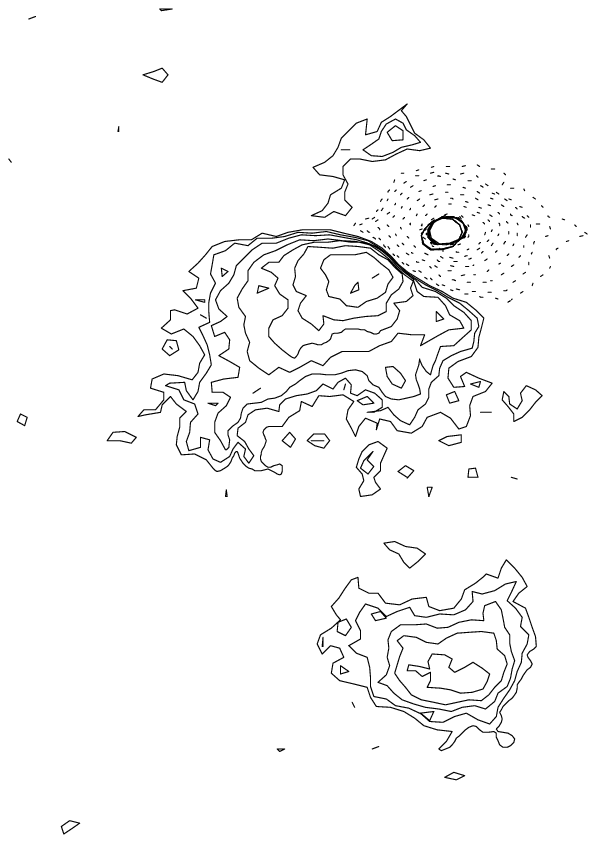}}
\put(10,-140){\framebox(255,205)}
\put(220,40){\textbf{H$_{2}$ - K}}
\thicklines
\put(220,20){\line(1,0){40.9}}
\put(235,5){5''}
\put(143,7){*}
\thinlines

\end{picture} 
\end{center}
\end{figure*}

\pagebreak

\begin{figure*}
\begin{center}
\begin{picture}(200,400)

\put(0,0){\includegraphics{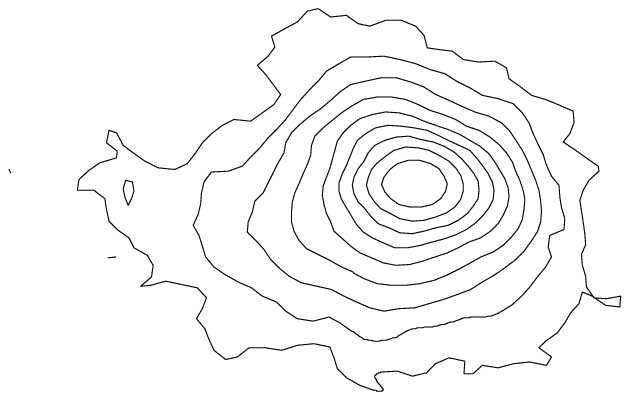}}
\put(-10,0){\framebox(230,200)}
\put(165,185){\textbf{L$^{\prime}$}}
\put(150,170){\textbf{(3.8~$\mu$m)}}
\thicklines
\put(160,150){\line(1,0){27.2}}
\put(165,135){2''}
\thinlines
\put(12,180){\textbf{04016+2610}}
\put(75,250){Figure 4}
\end{picture} 
\end{center}
\end{figure*}

\pagebreak

\begin{figure*}
\begin{center}
\begin{picture}(200,400)

\put(0,0){\includegraphics{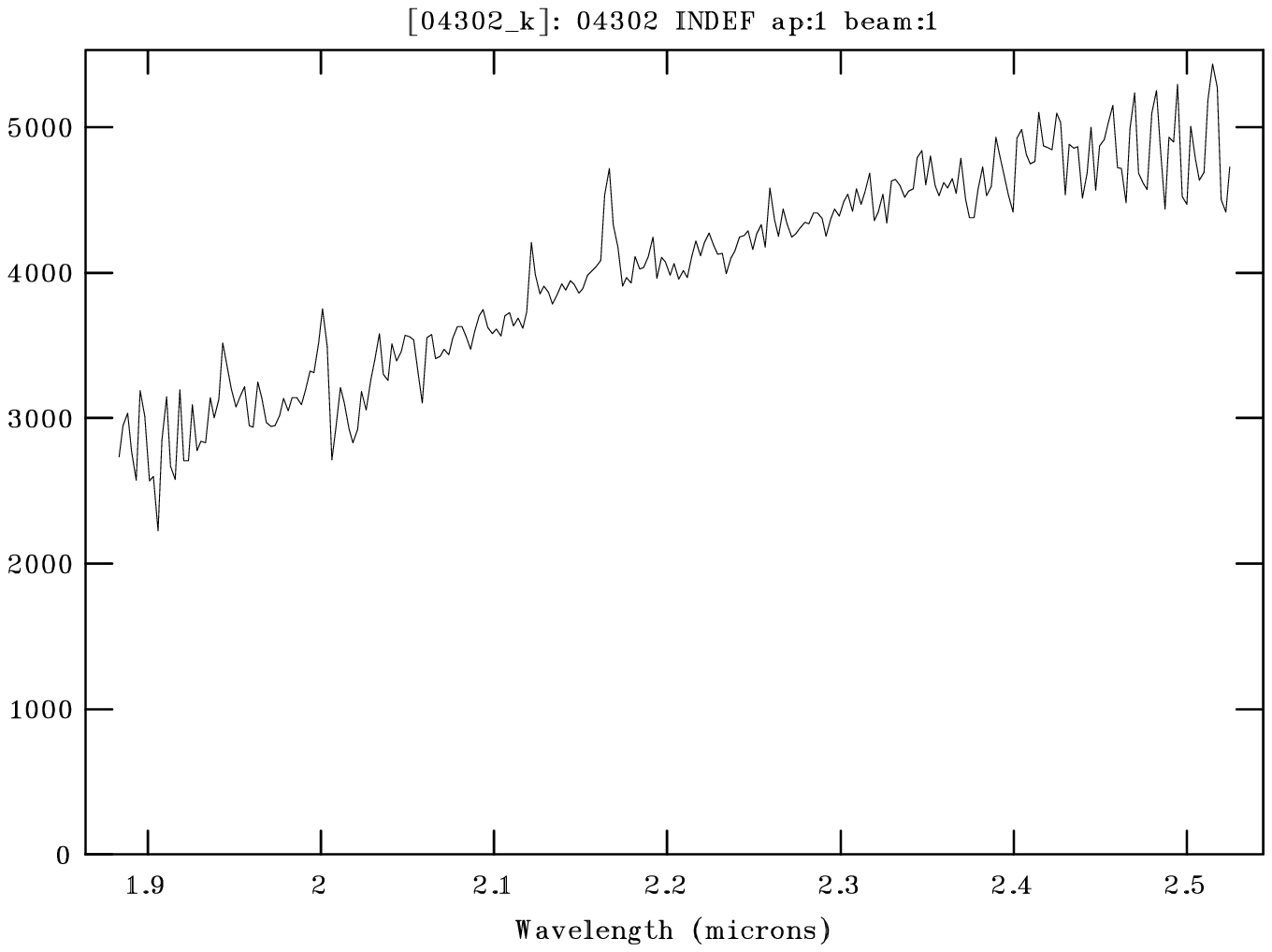}}
\put(-38,188){\textbf{04302+2247}}

\put(7,152){\line(0,1){5}}
\put(0,162){Br$\delta$}
\put(59,164){\line(0,1){5}}
\put(52,174){H$_{2}$}
\put(78,182){\line(0,1){5}}
\put(70,192){Br$\gamma$}

\put(-69,138){F$_{\lambda}$}
\put(-76,128){(DN)}
\put(75,250){Figure 5}
\end{picture} 
\end{center}
\end{figure*}

\end{document}